ITP Preprint Number NSF-ITP-9523# Weak Coupling Phase Diagram of the Two–Chain Hubbard Model

Leon Balents and Matthew P. A. Fisher
*Institute for Theoretical Physics, University of California, Santa Barbara, CA 93106-4030*(March 8, 1995)

We present a general method for determining the phase diagram of systems of a finite number of one dimensional Hubbard–like systems coupled by single–particle hopping with weak interactions. The technique is illustrated by detailed calculations for the two–chain Hubbard model, providing the first controlled results for arbitrary doping and inter-chain hopping. Of nine possible states which could occur in such a spin–1/2 ladder, we find seven at weak coupling. We discuss the conditions under which the model can be regarded as a one–dimensional analog of a superconductor.PACS: 71.26.+a, 74.20.Mn, 72.15.Nj

One dimensional (1d) electron systems provide an important testing ground for understanding electron correlation effects. Many methods have been applied to the problem of a single Hubbard chain, and there is general agreement that the system remains, for repulsive interactions, in a Luttinger liquid state with gapless spin and charge modes [1]. The 1d analog of a superconductor, a state with one gapless charge mode and dominant pairing (rather than charge density wave) correlations, does not arise in that case.

Two chain systems are interesting as a first step towards true 2d materials, and may be relevant for some experimental systems [2]. Moreover, on a ladder, statistics are more important, since particles can exchange without passing through one another. However, the theoretical situation in such models is much less clear [3–8]. Recent simulations suggest that states with dominant pairing correlations can indeed arise [9].

In this letter, we present a systematic weak coupling analysis of two Hubbard chains coupled by single particle hopping, $t_\perp$. Unlike previous work, our approach is a *controlled* renormalization group valid for small $U$ but for *arbitrary* inter–chain hopping and filling, $n$ [10]. The general methods described here may be applied to *any* system composed of a finite number of Hubbard–like chains with weak short range four–fermion interactions.

The possible phases of such models can be characterized by the number of charge and spin modes which are gapless at zero momentum. For an $N$-chain system the number of gapless charge modes can vary from zero to $N$, and likewise for spin. Remarkably, of the nine possible phases for two chains, seven are realized within the simple Hubbard model at weak coupling, reflecting the proliferation of marginal operators. Denoting a phase with $x$ gapless charge modes and $y$ gapless spin modes as CxSy, the small $U$ phase diagram for small fixed $t_\perp$ is shown in Fig. 1. Particularly noteworthy is the phase C1S0, present with purely *repulsive* interactions (positive $U$). This phase has a spin gap and a *single* gapless charge mode, and is thus the 1d analog of either a superconductor (SC) or charge density wave (CDW). We suggest two alternative physical criteria to distinguish the two possibilities. Preserving the resistance of the SC state to weak impurity scattering requires a slow decay of pairing correlations [11]. In particular, if the equal time pairing correlation function $\langle \Delta(x)\Delta^\dagger(0)\rangle \sim 1/|x|^\kappa$, where $\Delta = c_{1\uparrow}c_{2\downarrow}$, this requires $\kappa < \kappa_c = 1/3$. Alternatively, if an array of ladders is weakly coupled via Coulomb interactions and hopping, the SC state is more resilient with $\kappa_c \geq 1/2$. The traditional requirement [12] of "dominant SC correlations" gives $\kappa_c = 1$ (see below).

Interestingly, the spin-gapped C1S0 phase occurs in two different regimes (Fig. 1), one for doping, $\delta = 1-n$, away from half-filling, and the other when the Fermi energy coincides with a band edge, $k_{F1} = 0$. In the former case, pairing correlations develop upon doping the spin-gapped Mott insulator at half filling, $n = 1$, as in Anderson's original RVB picture for superconductivity in the Cuprates [13]. The critical doping $\delta_c$ at which C1S0 gives way to a gapless spin state, C2S1 and C2S2, is large for weak inter-chain hopping, decreasing from $\delta_c = 1$ for small $t_\perp$ to $\delta_c = 0$ as $t_\perp \to 2t$. Note that the phase C2S2 is the 1d analog of a Fermi liquid with all spin and charge modes gapless. The presence of the spin-gapped state (C1S0) near $k_{F1} = 0$, can be attributed to the coincidence of the Fermi energy with the Van Hove singularity at the 1d band edge.

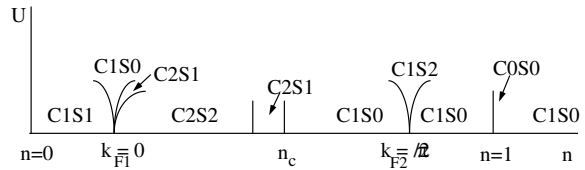

FIG. 1. Phase diagram at fixed $t_\perp < t$ (not to scale).

The two chain Hubbard model is described by the hamiltonian $H = H_0 + H_U$, with

$$H_0 = \sum_{x,\alpha} \left\{ -t\left(c^\dagger_{x,\alpha} c_{x+1,\alpha} + c \leftrightarrow d\right) - t_\perp c^\dagger_{x,\alpha} d_{x,\alpha} + h.c. \right\},$$
$$H_U = \sum_x U : \left( c^\dagger_{x,\uparrow} c_{x,\uparrow} c^\dagger_{x,\downarrow} c_{x,\downarrow} + c \leftrightarrow d \right) :, \qquad (1)$$

where $c$ $(c^\dagger)$ and $d$ $(d^\dagger)$ are fermion annihilation (creation) operators on the first and second chain, respectively, and $\alpha = \uparrow, \downarrow$ is a spin index. The parameters $t$ and $t_\perp$ are hopping matrix elements along and between the chains, and $U$ is an on–site Hubbard interaction. Eq.1 has the usual U(1)×SU(2) charge/spin symmetry.

For weak coupling it is natural to proceed by first diagonalizing the quadratic portion of the hamiltonian. This is achieved by canonically transforming to bonding and anti–bonding band operators: $\psi_{i\alpha} = (c_\alpha + (-1)^i d_\alpha)/\sqrt{2}$, with $i = 1, 2$. In momentum space $H_0$ becomes

$$H_0 = \sum_{i,\alpha} \int_{-\pi}^{\pi} \frac{dp}{2\pi} \epsilon_i(p) \psi^\dagger_{i\alpha}(p) \psi_{i\alpha}(p), \qquad (2)$$

where $\epsilon_1 = t_\perp - 2t\cos p$ and $\epsilon_2 = -t_\perp - 2t\cos p$. For $t_\perp > 2t$, the two bands are completely separated. At half–filling, the system is then a band insulator, and when doped becomes an ordinary spin-1/2 Luttinger liquid. For $t_\perp < 2t$, the bands overlap over some range of energies. When the Fermi level lies within this region, interaction effects must be re–examined in detail.

It is sufficient to consider the behavior of the system only near the two Fermi momenta $k_{Fi}$, defined by $\epsilon_i(k_{Fi}) = \mu$. The chemical potential, $\mu$, is fixed by the requirement $k_{F1} + k_{F2} = n\pi$, where $n$ is the particle number per site. The decomposition $\psi_{i\alpha} \approx \psi_{Ri\alpha} e^{ik_{Fi}x} + \psi_{Li\alpha} e^{-ik_{Fi}x}$ gives, up to a constant,

$$H_0 = \sum_{i,\alpha} \int dx\, v_i \left( \psi^\dagger_{Ri\alpha} i\partial_x \psi_{Ri\alpha} - \psi^\dagger_{Li\alpha} i\partial_x \psi_{Li\alpha} \right), \qquad (3)$$

where $v_i = 2t \sin k_{Fi}$. The allowed 4 fermi interactions are highly constrained by symmetry. In addition to U(2) invariance, these terms must be preserved by time reversal, parity, chain interchange, and spatial translation operations. At generic fillings, the two fermi momenta are incommensurate, and the symmetry under translations is effectively doubled into independent transformations in each band. To delineate the couplings in a physical way, we employ the notation of current algebra,

$$\begin{aligned}
J_{iR} &= \psi^\dagger_{Ri\alpha} \psi_{Ri\alpha}, & \mathbf{J}_{iR} &= \tfrac{1}{2} \psi^\dagger_{Ri\alpha} \boldsymbol{\sigma}_{\alpha\beta} \psi_{Ri\beta}, \\
L_R &= \psi^\dagger_{R1\alpha} \psi_{R2\alpha}, & \mathbf{L}_R &= \tfrac{1}{2} \psi^\dagger_{R1\alpha} \boldsymbol{\sigma}_{\alpha\beta} \psi_{R2\beta}, \\
M_{iR} &= -i\psi_{Ri\uparrow} \psi_{Ri\downarrow}, & N_{R\alpha\beta} &= \psi_{R1\alpha} \psi_{R2\beta}, \qquad (4)
\end{aligned}$$

where $\boldsymbol{\sigma}$ denotes Pauli matrices. Left–moving currents are defined analogously. There are eight allowed interactions connecting left and right movers for generic fillings, with Hamiltonian densities

$$\begin{aligned}
-\mathcal{H}^{(1)}_{\text{int}} &= \tilde{g}_{1\rho} J_{1R} J_{1L} + \tilde{g}_{2\rho} J_{2R} J_{2L} + \tilde{g}_{x\rho}(J_{1R} J_{2L} + J_{2R} J_{1L}) + \tilde{g}_{1\sigma} \mathbf{J}_{1R} \cdot \mathbf{J}_{1L} + \tilde{g}_{2\sigma} \mathbf{J}_{2R} \cdot \mathbf{J}_{2L} \\
&+ \tilde{g}_{x\sigma} (\mathbf{J}_{1R} \cdot \mathbf{J}_{2L} + \mathbf{J}_{2R} \cdot \mathbf{J}_{1L}) + \tilde{g}_{t\rho}(L_R L_L + L^\dagger_R L^\dagger_L) + \tilde{g}_{t\sigma}(\mathbf{L}_R \cdot \mathbf{L}_L + \mathbf{L}^\dagger_R \cdot \mathbf{L}^\dagger_L).
\end{aligned} \qquad (5)$$

Six additional interactions are completely chiral,

$$\begin{aligned}
-\mathcal{H}^{(2)}_{\text{int}} &= \tilde{\lambda}_{1\rho}(J^2_{1R} + J^2_{1L}) + \tilde{\lambda}_{2\rho}(J^2_{2R} + J^2_{2L}) + \tilde{\lambda}_{x\rho}(J_{1R} J_{2R} + J_{1L} J_{2L}) \\
&+ \tilde{\lambda}_{1\sigma}(\mathbf{J}_{1R} \cdot \mathbf{J}_{1R} + \mathbf{J}_{1L} \cdot \mathbf{J}_{1L}) + \tilde{\lambda}_{2\sigma}(\mathbf{J}_{2R} \cdot \mathbf{J}_{2R} + \mathbf{J}_{2L} \cdot \mathbf{J}_{2L}) + \tilde{\lambda}_{x\sigma}(\mathbf{J}_{1R} \cdot \mathbf{J}_{2R} + \mathbf{J}_{1L} \cdot \mathbf{J}_{2L}).
\end{aligned} \qquad (6)$$

The couplings in Eq.6 renormalize "velocities" of various charge and spin modes, and can be neglected to leading order in $U$ for what follows. Additional operators are needed to treat umklapp processes at special dopings:

$$\begin{aligned}
-\mathcal{H}^{(3)}_{\text{int}} &= \tilde{u}_1(M^\dagger_{1R} M_{1L} + M^\dagger_{1L} M_{1R}) + \tilde{u}_2(M^\dagger_{2R} M_{2L} + M^\dagger_{2L} M_{2R}) + \tilde{u}_x(M^\dagger_{1R} M_{2L} + M_{1R} M^\dagger_{2L} + M^\dagger_{2R} M_{1L} \\
&+ M_{2R} M^\dagger_{1L}) + \tilde{u}_{t1}(N^\dagger_{R\alpha\beta} N_{L\alpha\beta} + N_{R\alpha\beta} N^\dagger_{L\alpha\beta}) + \tilde{u}_{t2}(N^\dagger_{R\alpha\beta} N_{L\beta\alpha} + N_{R\alpha\beta} N^\dagger_{L\beta\alpha}).
\end{aligned} \qquad (7)$$

The single–band umklapp term, $\tilde{u}_i$, is non–zero only if $k_{Fi} = \pi/2$. At half-filling the three inter–band umklapp terms ($\tilde{u}_x, \tilde{u}_{t1}, \tilde{u}_{t2}$) are non–vanishing.

The Hubbard model values for the coupling constants, obtained from Eq.1, are shown in Table I. To analyze the behavior of the weakly interacting system, we employ the renormalization group (RG) approach. In the RG, short–wavelength modes are progressively eliminated in a systematic way, leading to differential equations for the renormalized coupling constants which describe the physics of the model at longer and longer length scales. The flow equations for this system in the absence of $u_i$ were first obtained in Ref. [14] using conventional diagrammatic methods. The full set of RG equations is more directly obtained using current algebra [15]. Away from half–filling, they are

$$\begin{aligned}
\dot{g}_{1\rho} &= \beta(g^2_{t\rho} + \tfrac{3}{16} g^2_{t\sigma}) - \alpha u^2_1, \\
\dot{g}_{2\rho} &= \alpha(g^2_{t\rho} + \tfrac{3}{16} g^2_{t\sigma}) - \beta u^2_2,
\end{aligned}$$

$$\dot{g}_{x\rho} = -(g_{t\rho}^2 + \frac{3}{16}g_{t\sigma}^2),$$

$$\dot{g}_{1\sigma} = -\alpha g_{1\sigma}^2 - \frac{\beta}{2}g_{t\sigma}^2 + 2\beta g_{t\rho}g_{t\sigma},$$

$$\dot{g}_{2\sigma} = -\beta g_{2\sigma}^2 - \frac{\alpha}{2}g_{t\sigma}^2 + 2\alpha g_{t\rho}g_{t\sigma},$$

$$\dot{g}_{x\sigma} = -g_{x\sigma}^2 - \frac{1}{2}g_{t\sigma}^2 - 2g_{t\rho}g_{t\sigma},$$

$$\dot{g}_{t\rho} = g_{0\rho}g_{t\rho} + \frac{3}{16}g_{0\sigma}g_{t\sigma},$$

$$\dot{g}_{t\sigma} = g_{0\sigma}g_{t\rho} + (g_{0\rho} - g_{0\sigma}/2 - 2g_{x\sigma})g_{t\sigma},$$

$$\dot{u}_1 = -2\alpha g_{1\rho}u_1, \qquad \dot{u}_2 = -2\beta g_{2\rho}u_2, \qquad (8)$$

where $\tilde{g}_i \equiv \pi(v_1 + v_2)g_i$, $\alpha \equiv (v_1 + v_2)/(2v_1)$, $\beta \equiv (v_1 + v_2)/(2v_2)$, $g_{0\rho} = \alpha g_{1\rho} + \beta g_{2\rho} - 2g_{x\rho}$, and $g_{0\sigma} = \alpha g_{1\sigma} + \beta g_{2\sigma} - 2g_{x\sigma}$. The dots indicate logarithmic derivatives with respect to the length scale, i.e. $\dot{g}_i \equiv \partial g_i/\partial \ell$, where $\ell = \ln L$.

Eqs.8 are valid until $\max\{g_i\} \sim O(1)$. To analyze them, we employ the following approach. Starting with the appropriate initial values (c.f. Table I), we integrate the equations numerically. If, as $\ell \to \infty$, all the couplings approach finite values, the procedure is controlled, since $\max\{g_i(\ell = \infty)\}$ becomes arbitrarily small as $U \to 0$. If any coupling diverges, we determine the asymptotic behavior of all the couplings with Eqs.8. As $U \to 0$, this asymptotic behavior is approached arbitrarily closely, as the divergence occurs for larger and larger $\ell$. While the fate of a particular choice of initial $\{g_i\}$ can only be determined numerically (due to the complexity of the equations), the possible asymptotic behaviors when some couplings diverge can be determined analytically.

To do so, we make the ansatz $g_i(\ell) = kg_{i0}/(1-k\ell)$, where $1/k$ is the scale at which the couplings diverge. Eqs.8 then reduce to a set of coupled quadratic equations for the $\{g_{i0}\}$. The search for appropriate solutions is considerably aided by the numerical integration of the flow equations. After locating a divergence (which fixes $k$), we plot $(1-k\ell)g_i$ versus $\ell$, from which $g_{i0}$ is extracted from the intercept with the line $\ell = 1/k$.

Applying this procedure for generic fillings with Hubbard initial values, we found three distinct phases (in the regime with both bands partially filled for $U = 0$). For $\alpha \gtrsim 4.8$, the flows are stable, with fixed point values $g_{i\sigma}^* = g_{x\sigma}^* = g_{t\sigma}^* = g_{t\rho}^* = 0$. When $4.3 \lesssim \alpha \lesssim 4.8$, the system is singly unstable, with $g_{2\sigma,0} = -1/\beta$, and all other $g_{i0} = 0$. For more comparable fermi velocities, $1 < \alpha \lesssim 4.3$, all the operators except $g_{x\sigma0}^* = 0$ diverge, but in such a way that $\alpha g_{1\sigma0} = \beta g_{2\sigma0}$ and $g_{t\sigma0} = -4g_{t\rho0}$. The behavior for $1/2 < \alpha < 1$ is obtained by interchanging band indices in all quantities.

The physics of these phases is elucidated through the use of abelian bosonization [12,10]. With the convention $\psi_{R/Li\alpha} \propto \exp(i\sqrt{4\pi}\phi_{R/Li\alpha})$, dual canonical Bose fields may be defined as $\phi_{i\alpha} = \phi_{Ri\alpha} + \phi_{Li\alpha}$ and $\theta_{i\alpha} = \phi_{Ri\alpha} - \phi_{Li\alpha}$. They satisfy $[\phi(x), \theta(y)] = -i\mathrm{sgn}(x-y)/2$. A further canonical transformation to $(\phi, \theta)_{i\rho} = [(\phi, \theta)_{i\uparrow} + (\phi, \theta)_{i\downarrow}]/\sqrt{2}$ and $(\phi, \theta)_{i\sigma} = [(\phi, \theta)_{i\uparrow} - (\phi, \theta)_{i\downarrow}]/\sqrt{2}$ yields the spin–charge separated Euclidean action

$$S_0 = \sum_{i\nu}\int_{x,\tau} \frac{v_i}{2}\left[(\partial_x\phi_{i\nu})^2 + (\partial_x\theta_{i\nu})^2\right] + i\partial_x\theta_{i\nu}\partial_\tau\phi_{i\nu}, \quad (9)$$

where $\nu = \rho, \sigma$. Using the scheme discussed in the introduction, the non–interacting system with both bands occupied (Eq.9) is classified as C2S2. The large $\alpha$ phase found above is also of C2S2 type, though it contains the additional (marginal) couplings $g_{i\rho}, g_{x\rho}, \lambda_{i\rho}, \lambda_{x\rho}, \lambda_{i\sigma}$, and $\lambda_{x\sigma}$, which makes the behavior highly non–universal [16].

In the intermediate state ($4.3 \lesssim \alpha \lesssim 4.8$), $g_{2\sigma}$ becomes large and negative. Using bosonization, this interaction (neglecting unimportant gradient terms) is

$$S_{2\sigma} \propto \tilde{g}_{2\sigma}\int_{x,\tau} M^2\cos(\sqrt{8\pi}\theta_{2\sigma}), \quad (10)$$

where the coefficient $M$ is cut–off dependent. In the scaling limit, it is appropriate to expand the cosine and obtain a true mass $M$ for $\theta_{2\sigma}$. The resulting phase is therefore C2S1. For $\alpha \lesssim 4.3$, an analogous cosine appears in the $1\sigma$ sector, and the asymptotic divergence of $g_{t\rho}$ and $g_{t\sigma}$ is such that the inter–band hopping terms sum to

$$S_t \propto \tilde{g}_{t\rho}\int_{x,\tau}\cos(\sqrt{4\pi}\phi_{-\rho})\cos(\sqrt{2\pi}\theta_{1\sigma})\cos(\sqrt{2\pi}\theta_{2\sigma}), \quad (11)$$

where $\phi_{\pm\rho} \equiv (\phi_{1\rho} \pm \phi_{2\rho})/\sqrt{2}$. It is natural to assign masses to $\theta_{1\sigma}$ and $\theta_{2\sigma}$, after which Eq.11 acts as a mass for $\phi_{-\rho}$. The resulting state has only a single ungapped charge mode (with dual fields $\theta_{+\rho}, \phi_{+\rho}$), and will be labeled C1S0 [10]. It may be characterized by a single dimensionless stiffness $K_{+\rho}$ and a velocity $v_{+\rho}$, such that the effective action for $\theta_{+\rho}$ is

$$S_{+\rho} = \frac{K_{+\rho}}{2}\int_{x,\tau}\left\{v_{+\rho}(\partial_x\theta_{+\rho})^2 + v_{+\rho}^{-1}(\partial_\tau\theta_{+\rho})^2\right\}, \quad (12)$$

with a similar dual ($K_{+\rho} \to K_{+\rho}^{-1}$) form for $\phi_{+\rho}$. Eq.12 interpolates smoothly between a CDW (at large $K_{+\rho}$) and a superconductor (at small $K_{+\rho}$). The pairing exponent $\kappa = K_{+\rho}/2$, while the CDW–like correlator $\langle n^2(x)n^2(0)\rangle_c \sim \cos[2(k_{F1} + k_{F2})x]/x^{2/K_{+\rho}}$, plus power laws $(1/x^2)$ at $k = 0$. We are unable to determine $K_{+\rho}$ and $v_{+\rho}$ in a controlled way, because they depend on the entire crossover from the non–interacting state to the C1S0 fixed point. However, heuristic calculations suggest that superconducting fluctuations ($K_{+\rho}^{-1}$) increase with increasing $v_1 - v_2$ and decrease with interactions. Interestingly, the usual power law term at $2k_F$ is missing from the density–density correlation function in the C1S0 phase, as noted by Nagaosa [8] in a similar model, due to strong fluctuations of the $\theta_{-\rho}$ and $\phi_{i\sigma}$ fields.

It remains to discuss the behavior at several "special" points in the phase diagram. When $k_{F2} = \pi/2$, umklapp processes imply $u_2 \neq 0$. For $t_\perp > t$, this occurs with band 1 empty, and one gets the usual C0S1 spin density wave. For $t_\perp < t$, we must consider inter–band coupling via Eqs.8. For almost all ratios of the velocities, we find that a charge gap develops in band 2, simultaneously suppressing the other potential instabilities, leading to a C1S2 phase. Surprisingly, over the narrow range $0.6 \lesssim \beta \lesssim 0.85$, $u_2$ renormalizes to zero, yielding instead the C1S0 state. The behavior at half–filling is more difficult to obtain, because it requires the inclusion of the $u_x$, $u_{t1}$, and $u_{t2}$ operators in Eq.7. The RG equations in this case are too complicated to reproduce here [15]. Their analysis indicates a completely gapped (C0S0) phase, as suggested by a large $U$ picture of coupled antiferromagnetic Heisenberg chains [15].

The final remaining special point occurs when the fermi level lies precisely at the bottom of band 1. It is outside the scope of conventional RGs, because the dispersion in band 1 is quadratic, with the hamiltonian

$$H_1 = -\int_x \frac{1}{2\tilde{m}^\epsilon} \psi_1^\dagger \partial_x^{1+\epsilon} \psi_1, \quad (13)$$

where $\epsilon = 1$ for the quadratic band, but must be taken as a small parameter to control the perturbative treatment. The allowed couplings are $g_{2\rho}$, $g_{2\sigma}$, $\lambda_{2\rho}$, and the four interband terms

$$-\mathcal{H}_{1-2} = \tilde{u}_\rho(J_{R2} + J_{L2})J_1 + \tilde{u}_\sigma(\mathbf{J}_{R2} + \mathbf{J}_{L2}) \cdot \mathbf{J}_1$$
$$- \frac{\tilde{v}}{2} J_1^2 - \tilde{u}_t(\psi_{R2\alpha}^\dagger \psi_{L2\beta}^\dagger \psi_{1\beta} \psi_{1\alpha} + \text{h.c.}), \quad (14)$$

where $J_1 = \psi_{1\alpha}^\dagger \psi_{1\alpha}$ and $\mathbf{J}_1 = \psi_{1\alpha}^\dagger \boldsymbol{\sigma}_{\alpha\beta} \psi_{1\beta}/2$. The RG equations in this case are [15]

$$\dot{g}_{2\rho} = u_t^2/2, \qquad \dot{g}_{2\sigma} = -g_\sigma^2 - 2u_t^2,$$
$$\dot{u}_\rho = \gamma m^{-\epsilon} u_t^2, \qquad \dot{u}_\sigma = -\gamma u_\sigma^2,$$
$$\dot{v} = \epsilon v - 2v^2 - u_t^2,$$
$$\dot{u}_t = (\epsilon/2 - 2v - 4\gamma u_\rho - 3g_\sigma/4 + g_\rho)u_t, \quad (15)$$

where $m = 2\tilde{m}^\epsilon$, $\gamma = 1/(1+m^{-\epsilon})$, $u_t = m^{\epsilon/2} \tilde{u}_t$, $v = m^\epsilon \tilde{v}/(4\pi)$, $(g_2, u)_{\rho,\sigma} = (\tilde{g}_2, \tilde{u})_{\rho,\sigma}/(2\pi v_2)$ (we have taken a momentum cut-off of 1). Because of the relative simplicity of Eqs.15, we have been able to analytically show an instability for Hubbard initial conditions [15]. Analysis of the flow asymptotics indicates that at this special point, fluctuation induced attractive interactions $v < 0$ in band 1 populate it with spinless bound pairs. Simultaneously, $g_{2\sigma} < 0$ creates a spin gap in band 2, and $u_t$ Josephson couples the two bands, gapping the out of phase charge mode to leave a C1S0 phase. This RG analysis holds for $U \ll \epsilon \ll 1$, but we expect the tendency to pairing to *increase* with $\epsilon$, due to the increased density of states for inter–band scattering. Physically, the presence of the C1S0 phase at $k_{F1} = 0$ can be attributed to the Van Hove singularity at the band edge.

The above results at special fillings are valid only at isolated points for infinitesimal $U$. For finite $U$, the RG suggests that the regions of attraction of these phases widen into fans of width $\delta t_\perp \sim \exp(-ct/U)$, where $c$ is a constant. The non–uniformity of Eqs.15 allows for additional structure within the fan. In particular, because the instability is driven by the $g_{2\sigma}$ term, we expect a narrow intermediate wedge of C2S1, as shown in Fig.1.

It will be interesting to generalize the above calculations to three-chain systems (N=3), to help clarify which features are particular to even N.

We are grateful to A. Ludwig and especially D. Scalapino for innumerable fruitful discussions. This work has been supported by the National Science Foundation under grants No. PHY89-04035 and No. DMR-9400142.

TABLE I. Hubbard model coupling constants. The g–ology notation is given for comparison with Ref.14.

| Coupling | g–ology | Hubbard value |
|---|---|---|
| $\tilde{g}_{1\rho}$ | $g^1_{AAAA}/2 - g^2_{AAAA}$ | $-U/4$ |
| $\tilde{g}_{2\rho}$ | $g^1_{BBBB}/2 - g^2_{BBBB}$ | $-U/4$ |
| $\tilde{g}_{x\rho}$ | $g^1_{ABAB}/2 - g^2_{ABBA}$ | $-U/4$ |
| $\tilde{g}_{t\rho}$ | $g^1_{AABB}/2 - g^2_{AABB}$ | $-U/4$ |
| $\tilde{g}_{1\sigma}$ | $2g^1_{AAAA}$ | $U$ |
| $\tilde{g}_{2\sigma}$ | $2g^1_{BBBB}$ | $U$ |
| $\tilde{g}_{x\sigma}$ | $2g^1_{ABAB}$ | $U$ |
| $\tilde{g}_{t\sigma}$ | $2g^1_{AABB}$ | $U$ |